\documentclass[
	a4paper, 
	10pt, 
	unnumberedsections, 
	twoside, 
]{LTJournalArticle}
\usepackage[dvipsnames]{xcolor}

\title{The Impact of Data Compression in Real-Time and Historical Data Acquisition Systems on the Accuracy of Analytical Solutions} 

\author{Reham Faqehi\textsuperscript{1} \\ \small\href{mailto:reham.faqihi@aramco.com}{reham.faqihi@aramco.com} 
   \and Haya Alhuraib\textsuperscript{1} \\\small\href{mailto:haya.alhuraib.1@aramco.com}{haya.alhuraib.1@aramco.com}
   \and Hamad Saiari\textsuperscript{1} \\ \small\href{mailto:hamad.alsaiari@aramco.com}{hamad.alsaiari@aramco.com}
   \and Zyad Bamigdad\textsuperscript{1} \\ \small\href{mailto:zyad.bamigdad@aramco.com}{zyad.bamigdad@aramco.com}}
\date{\footnotesize\textsuperscript{\textbf{1}}Process \& Control Systems Department, Saudi Aramco}

\renewcommand{\maketitlehookd}{%
	\begin{abstract}
		\noindent In industrial and IoT environments, massive amounts of real-time and historical process data are continuously generated and archived. With sensors and devices capturing every operational detail, the volume of time-series data has become a critical challenge for storage and processing systems. Efficient data management is essential to ensure scalability, cost-effectiveness, and timely analytics. To minimize storage expenses and optimize performance, data compression algorithms are frequently utilized in data historians and acquisition systems. However, compression comes with trade-offs that may compromise the accuracy and reliability of engineering analytics that depend on this compressed data. Understanding these trade-offs is essential for developing data strategies that support both operational efficiency and accurate, reliable analytics. This paper assesses the relation of common compression mechanisms used in real-time and historical data systems and the accuracy of analytical solutions, including statistical analysis, anomaly detection, and machine learning models. Through theoretical analysis, simulated signal compression, and empirical assessment, we illustrate that excessive compression can lose critical patterns, skew statistical measures, and diminish predictive accuracy. The study suggests optimum methods and best practices for striking a compromise between analytical integrity and compression efficiency.
	\end{abstract}
}

\begin{document}

\maketitle 


\section{Introduction}

Modern industrial systems depend significantly on continuous monitoring and control through real-time and historical data acquisition platforms. These platforms serve as the backbone for operational decision-making, aggregating vast amounts of data from diverse sources, including sensors, instruments, and control systems. As industrial processes grow in complexity and automation, the scale and granularity of collected data increase exponentially, posing significant challenges in storage management, retrieval latency, and system scalability.

The increasing deployment of industrial internet of things (IIoT) devices and pervasive sensing technologies has further intensified the data deluge, with high-frequency measurements and long-term archival requirements contributing to vast repositories of time-series data. In such environments, efficient data management becomes critical not only for operational efficiency but also for enabling advanced analytics, predictive maintenance, and optimization strategies that rely on timely and accurate information. The rapid proliferation of the Industrial Internet of Things (IIoT) and pervasive sensing technologies has further intensified this data deluge. High-frequency measurements—such as vibration data from rotating machinery, temperature fluctuations in chemical processes, or pressure readings in pipeline networks—generate massive volumes of time-series data that must be stored for regulatory compliance, long-term trend analysis, and retrospective investigations. Furthermore, industries such as manufacturing, energy, and utilities require multi-year data retention for auditing, predictive maintenance, and process optimization, leading to petabyte-scale historical archives. In such data-intensive environments, efficient storage and retrieval mechanisms are critical not only for operational efficiency but also for enabling advanced analytics, machine learning (ML)-driven predictive maintenance, and real-time optimization strategies. These applications rely on high-fidelity, temporally consistent data to detect subtle anomalies, forecast equipment failures, and optimize production throughput. However, the sheer volume of raw, uncompressed data strains storage infrastructure, increases costs, and slows down query performance. 

To handle the extensive amount of data generated, particularly in historical archives, compression techniques are implemented to minimize redundancy and improve storage and processing efficiency [1]. Although compression facilitates large-scale data management, its impact on data integrity raises concerns for analytics. Analytical applications, including Key Performance Indicators (KPIs), predictive maintenance, and anomaly detection relay on fine-grained temporal patterns, trends, and events transitions that may be lost by applying data compression. Therefore, striking a balance between compression efficiency and analytical accuracy has emerged as a pressing challenge in industrial data management. A deeper understanding of how various compression techniques affect analytical outcomes is essential for developing data strategies that optimize both resource usage and insight generation. This study examines, from a conceptual and empirical standpoint, how to balance between data compression and analytics accuracy. By examining real-world case studies from process manufacturing and energy monitoring, we assess the performance of various compression algorithms in preserving critical features for anomaly detection, trend analysis, and predictive modelling. Furthermore, we propose a framework for selecting compression strategies based on domain-specific requirements, ensuring that storage efficiency does not come at the expense of actionable insights. The primary contributions of this paper can be summarized as follows:

\begin{enumerate}
	\item How do different compression techniques impact the accuracy of industrial analytics?
	\item What are the trade-offs between compression ratios and the preservation of diagnostically relevant data features?
	\item Can hybrid compression approaches optimize both storage costs and analytical utility?
	\item How can domain knowledge (e.g., critical sampling rates for specific sensors) guide compression parameter selection?
\end{enumerate}

By addressing these points, this study aims to provide actionable guidelines for industrial data architects and operations teams, enabling them to implement compression strategies that align with both economic constraints and analytical requirements.


\section{Methodology}

This research follows a mixed-method approach that integrates literature review, simulation, and comparison analysis.

\subsection{Literature Review}

The increasing reliance on real-time and historical data in industrial systems has led to extensive research on data compression techniques and their impact on analytics. Lossless compression techniques have been explored in several studies. For instance, the work in [1] presents a lossless compression algorithm for wind plant data, achieving a compression ratio of up to 10:1. The algorithm uses a combination of techniques, including run-length encoding and Huffman coding, to achieve high compression ratios. Similarly, [2] proposes a singular value decomposition (SVD) based compression method for smart distribution systems, resulting in a compression ratio of up to 20:1. The SVD technique is used to reduce the dimensionality of the data, making it more compressible.

Historical data compression is crucial for power dispatch SCADA systems. The research in [3] investigates the compression and storage strategy for historical data in power dispatch SCADA systems, highlighting the importance of efficient data compression for reducing storage costs. The study proposes a compression algorithm that uses a combination of techniques, including data aggregation and compression, to achieve high compression ratios. Another study [4] applies tensor Tucker decomposition to compress historical multi-station SCADA data in distribution management systems, achieving a compression ratio of up to 30:1. The tensor Tucker decomposition technique is used to reduce the dimensionality of the data, making it more compressible.

\begin{table*} 
	\caption{Meaning of Different Threshold Levels}
	\centering 
	\begin{tabular}{L{0.2\linewidth} L{0.2\linewidth} L{0.40\linewidth}} 
		\toprule
		Threshold Level & Definition & Rationale \\
		\midrule
		Conservative & 10\% of signal standard deviation & Preserves fine details for anomaly detection or critical analytics. \\
		Moderate & 2\% of signal range & Balanced trade-off used in many historian default configurations.\\
		Aggressive & 25\% of signal standard deviation & Suitable for visualization or archival data where small deviations are tolerable.\\
		\bottomrule
	\end{tabular}
\end{table*}

\begin{table*} 
	\centering 
	\begin{tabular}{L{0.2\linewidth} L{0.4\linewidth}} 
		\toprule
		Ratio Value & Interpretation \\
		\midrule
		Close to 0 & Signal is mostly stable or flat \\
		Moderate (~0.1–0.3) & Signal has moderate fluctuations \\
		Close to 1 & Signal varies wildly throughout its range \\
		\bottomrule
	\end{tabular}
    \caption{Interpretation of Normalized Fluctuation Index}

\end{table*}

Wavelet-based compression techniques have been widely used in power systems. The work in [5] presents a novel data compression technique using adaptive fuzzy logic for power waveforms, resulting in a compression ratio of up to 15:1. The adaptive fuzzy logic technique is used to select the most suitable wavelet basis for compression, achieving high compression ratios. Similarly, [6] proposes a Huffman coding approach with wavelet transform enhancement for disturbance data compression in power systems, achieving a compression ratio of up to 20:1. The wavelet transform is used to reduce the dimensionality of the data, making it more compressible.

Other data compression techniques have also been investigated. For example, [7] presents a real-time power-quality monitoring system that uses a hybrid sinusoidal and lifting wavelet compression algorithm, achieving a compression ratio of up to 10:1. The hybrid algorithm combines the advantages of sinusoidal and lifting wavelet transforms to achieve high compression ratios. Another study [8] proposes a high-efficient compression method for power quality applications, achieving a compression ratio of up to 20:1. The method uses a combination of techniques, including data aggregation and compression, to achieve high compression ratios.

Data compression techniques have also been applied in smart grid applications. For instance, [9] presents a wavelet-based data compression technique for smart grids, achieving a compression ratio of up to 15:1. The wavelet transform is used to reduce the dimensionality of the data, making it more compressible. Another study [10] proposes a real-time data compression and adapted protocol technique for wide-area measurement systems (WAMS), achieving a compression ratio of up to 20:1. The technique uses a combination of data compression and protocol adaptation to achieve high compression ratios.

\begin{table*}
    \centering
    
    \begin{tabular}{L{0.2\linewidth} L{0.05\linewidth} L{0.05\linewidth} L{0.2\linewidth} L{0.1\linewidth} L{0.1\linewidth}}
         \toprule
         Signal Type & Std Dev & Signal Range & Threshold  & Compressed Points & Compression Rate (\%)\\
         \midrule
         Temperature & 0.96 & 4.80 & 0.096 (Moderate Compression) & 32 & 96.8\\
         Vibration & 3.55 & 11.77 & 0.355 (Conservative compression) & 331 & 66.9\\
    \end{tabular}
    \caption{Effects of Compression on Two Types of Signals}
\end{table*}

The reviewed studies collectively emphasize that:
\begin{itemize}
	\item Lossy compression improves storage efficiency but risks analytical accuracy, particularly for high-frequency and anomaly-detection applications.
	\item Hybrid and adaptive methods (e.g., wavelet transforms, tiered compression) offer promising trade-offs.
	\item Domain-specific tuning is crucial, as different industrial applications have varying sensitivity to data distortion.
\end{itemize}

While existing studies have significantly advanced our understanding of industrial data compression and its analytical implications, several critical gaps remain unaddressed in the literature. (1) there are no standardized metrics to evaluate compression's impact on analytical accuracy across different industrial applications; (2) the interaction between compression techniques and advanced machine learning models remains underexplored, particularly for real-time adaptive systems; (3) most studies neglect human factors and operational decision-making needs in compression design; and (4) comprehensive cost-benefit analyses considering long-term system-wide impacts are lacking. These limitations hinder the development of optimal compression strategies that balance storage efficiency with analytical fidelity in Industry 4.0 environments.

\subsection{Simulated Signal Experiment}

A synthetic time-series signal mimicking sensor behavior was generated, and different compression thresholds were applied utilizing a swinging-door algorithm to determine meaningful compression thresholds, we adopted a heuristic-based approach inspired by standard practices in time-series data analysis and industrial data historians (e.g., OSIsoft PI System). Threshold levels were selected as follows:

These thresholds are derived from literature and practical recommendations, including [1] and Bagnall et al. [12], as well as empirical tuning commonly applied in data compression systems.

An adaptive rule was developed to choose the optimum compression threshold based on the relative fluctuation of the signal.

\[std\_dev/ signal\_range = Normalized\ Fluctuation\ Index\]

Original and compressed datasets were examined to assess the impact on:

\begin{itemize}
	\item Trend representation
	\item Statistical measures (mean, standard deviation)
	\item Anomaly detection
\end{itemize}

\subsection{Analytical Accuracy Evaluation}

We trained multiple univariate models on both original and compressed datasets to analyze and compare prediction accuracy (MSE, MAE) to understand the impact of implementing such compression algorithms.

\section{Results}

\subsection{Compression Behavior and Signal Type Sensitivity}
Compression effectiveness depends strongly on the signal’s variability and shape. Using the swinging door algorithm with auto-suggested thresholds, we tested both a temperature and a vibration signal, each with 1000 points. The results are summarized below:

\begin{itemize}
	\item \textbf{Vibration} signals retained more data (low compression) due to its high-frequency fluctuations, which led to more turning points.
	\item \textbf{Temperature} signals were compressed more aggressively (high compression) with minimal loss of information.
\end{itemize}

These findings highlight the importance of applying context-specific thresholds when designing compression strategies. Figure 1 \& 2 illustrates how intermediate data points are discarded, while turning points are retained to preserve trend shape.

\begin{figure}
    \centering
    \includegraphics[width=\linewidth]{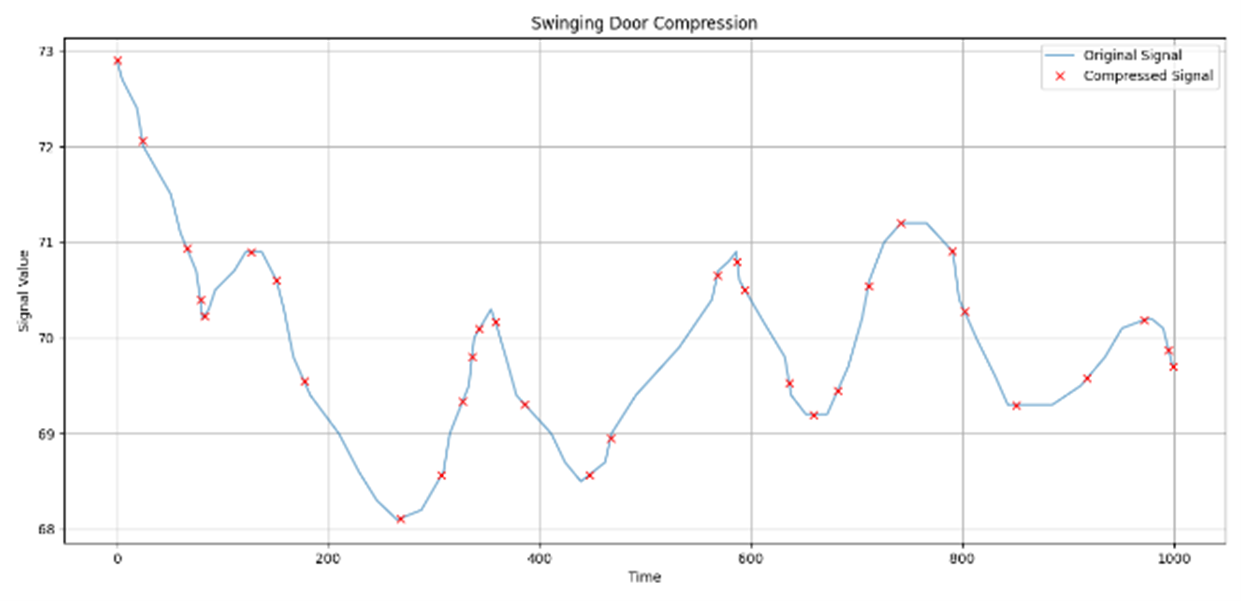}
    \caption{Compression Applied to Temperature Signal}
\end{figure}

\begin{figure}
    \centering
    \includegraphics[width=\linewidth]{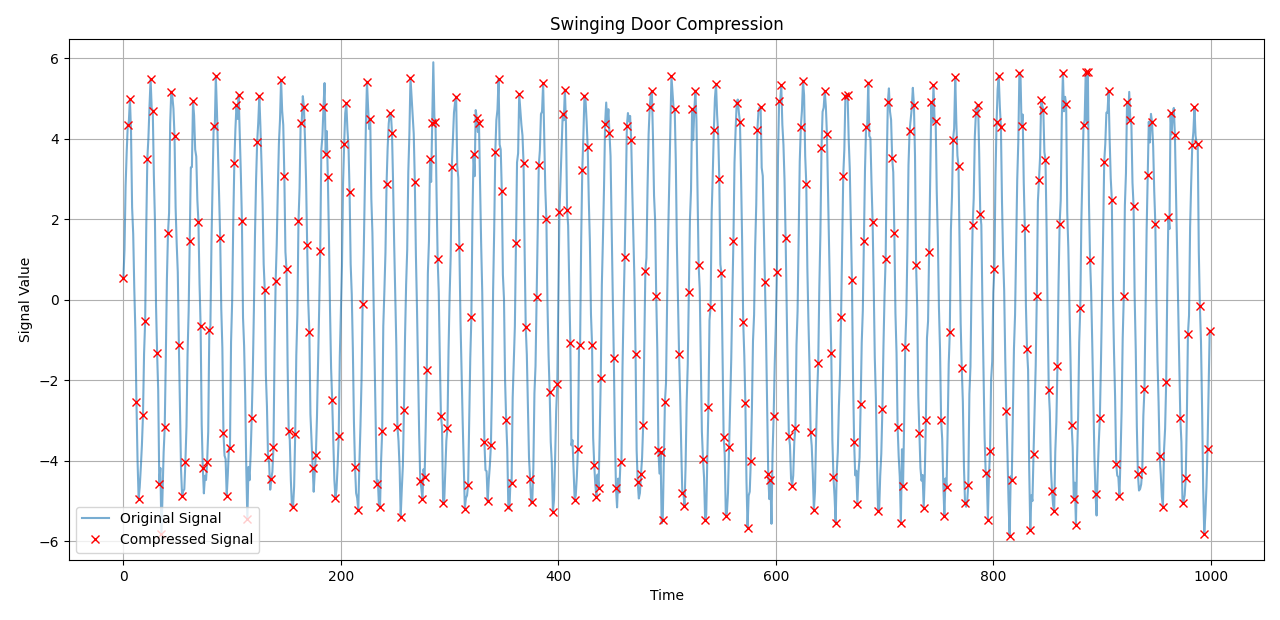}
    \caption{Compression Applied to Vibration Signal}
\end{figure}

In shape-preserving algorithms such as the swinging door method, data points that fall along a consistent slope are typically discarded, while turning points are retained to preserve signal integrity. As such, signals with frequent fluctuations or high-frequency noise tend to yield lower compression ratios, since more turning points must be preserved.

Additionally, the sensitivity and criticality of the parameter being monitored influence compression strategy. For instance, safety-critical or rapidly changing parameters may require more conservative thresholds to avoid loss of important dynamics.

The expected behavior and physical constraints of the signal should also inform compression configuration. For example, a boiler temperature signal is not expected to rise from 0°C to 100°C within two seconds, and such a transition may indicate an anomaly or sensor error. Incorporating these expectations allows tuning of the compression algorithm (e.g., setting appropriate thresholds or time windows) to ensure meaningful data reduction without compromising analytical fidelity.

\subsection{Data Reduction and Storage Efficiency}
Through multiple experimental trials using the swinging door algorithm, we applied various thresholds to a synthetic signal of 1000 points. To better reflect real-world distortion, RMSE was calculated using interpolation of the compressed signal against the full original time series. The following table summarizes the outcomes:

\begin{table*} 
	\centering 
    \caption{RMSE of Different Threshold Values}
	\begin{tabular}{C{0.15\linewidth} C{0.15\linewidth} C{0.15\linewidth} C{0.15\linewidth} C{0.15\linewidth}} 
        \cmidrule(r){1-2}
        
		\multicolumn{1}{l}{Original Data Points} & 1000 \\
        \multicolumn{1}{l}{Standard Deviation} & 7.13\\
		\multicolumn{1}{l}{Signal Range} & 22.22\\

		\midrule
		Threshold Value & Compressed Points & Compression Ratio & Data Reduction (\%) & RMSE (Interpolated) \\
		\midrule
        0.2 & 386 & 0.386 & 61.4 & 0.432 \\
		0.3 & 330 & 0.33 & 67 & 0.452 \\
		0.5 & 224 & 0.224 & 77.6 & 0.493 \\
		0.7 & 173 & 0.173 & 82.7 & 0.546 \\
		1 & 120 & 0.12 & 88 & 0.741 \\
		1.05 & 117 & 0.117 & 88.3 & 0.750 \\
		\bottomrule
	\end{tabular}
    
\end{table*}

\begin{table*} 
	\centering 
	\begin{tabular}{C{0.3\linewidth} C{0.1\linewidth} C{0.1\linewidth} C{0.1\linewidth} C{0.1\linewidth} C{0.1\linewidth}} 

		\toprule
		Threshold Type & Threshold Value & Compressed Points & Compression Ratio & Data Reduction (\%) & RMSE (Interpolated) \\
		\midrule
        Conservative (10\% of std) & 0.713 & 169 & 0.169 & 83.1 & 0.555 \\
		Moderate (2\% of range) & 0.444 & 257 & 0.257 & 74.3 & 0.474 \\
		Aggressive (25\% of std) & 1.783 & 80 & 0.08 & 92 & 1.41 \\
		
        \bottomrule
		\multicolumn{1}{c}{Suggested} & 0.713 \\
		\cmidrule(r){1-2}
	\end{tabular}
    \caption{RMSE of Different Threshold Values.}
\end{table*}

\begin{figure}
    \centering
    \includegraphics[width=\linewidth]{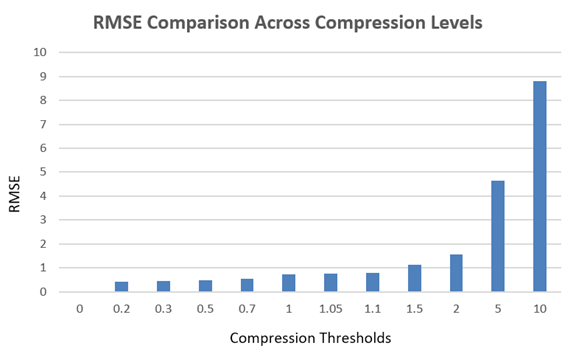}
    \caption{Compression Threshold vs RMSE}
\end{figure}

Lower threshold value give lower data reduction in the storage. But, better interpolation to get the data and low RMSE as the signal shape will be more protected.

In the other hand, higher threshold value give higher data reduction in the storage size. But, worse in interpolation to get the data and high RMSE as the signal shape will be more destroyed.

Moreover, using the swinging door algorithm with auto-suggested thresholds level algorithm. The optimum threshold was chosen based on adaptive rule that chooses the best compression threshold based on the relative fluctuation of the signal.

\begin{table*} 
	\caption{RMSE of Different Threshold Values.}
	\centering 
	\begin{tabular}{C{0.3\linewidth} C{0.1\linewidth} C{0.1\linewidth} C{0.1\linewidth} C{0.1\linewidth} C{0.1\linewidth}} 

		\toprule
		Threshold Type & Threshold Value & Compressed Points & Compression Ratio & Data Reduction (\%) & RMSE (Interpolated) \\
		\midrule
        Conservative (10\% of std) & 0.713 & 169 & 0.169 & 83.1 & 0.555 \\
		Moderate (2\% of range) & 0.444 & 257 & 0.257 & 74.3 & 0.474 \\
		Aggressive (25\% of std) & 1.783 & 80 & 0.08 & 92 & 1.41 \\
		
        \bottomrule
		\multicolumn{1}{c}{Suggested} & 0.713 \\
		\cmidrule(r){1-2}
	\end{tabular}
\end{table*}

To adapt the compression threshold to signal variability, we used the ratio of standard deviation to signal range as a heuristic. Based on this ratio, the algorithm selects a conservative, moderate, or aggressive threshold, ensuring the compression strategy aligns with the signal’s dynamic behavior. The threshold boundaries (e.g., 0.1 and 0.3) were selected based on the nature of the signal types used in our study. These values can be adjusted in practice to suit different signal characteristics.

As a result the optimum Compression of this synthetic signal is at 0.713 with reduction of 83.1\% and RMSE of 0.555

\begin{figure}
    \centering
    \includegraphics[width=\linewidth]{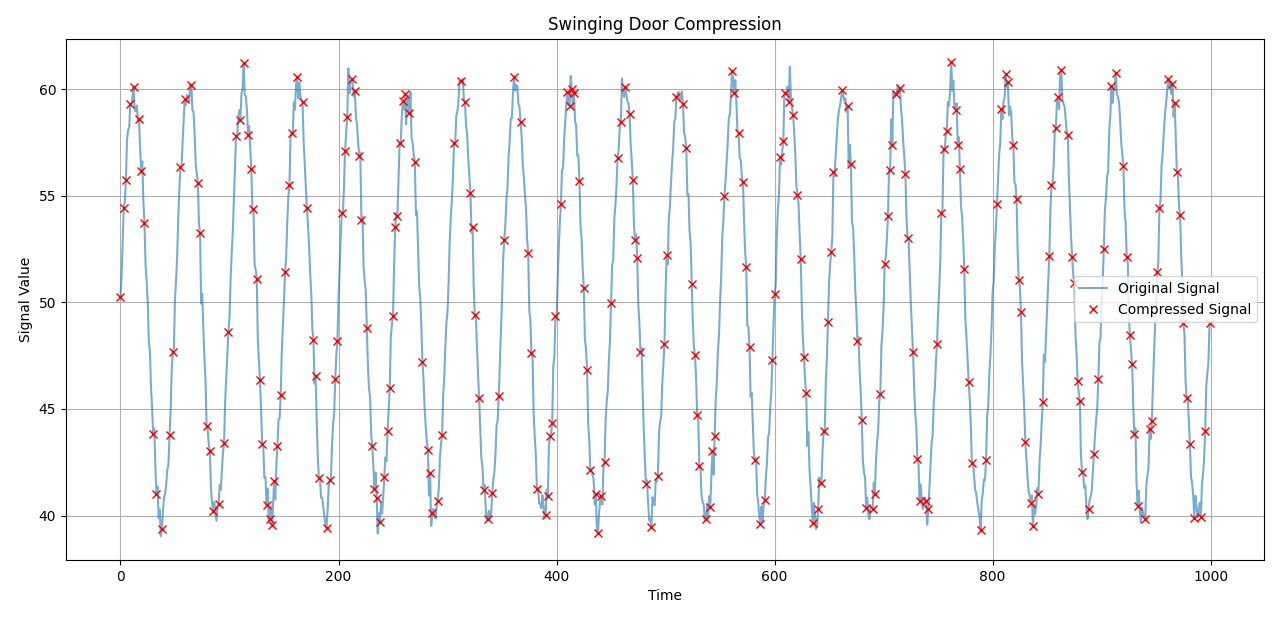}
    \includegraphics[width=\linewidth]{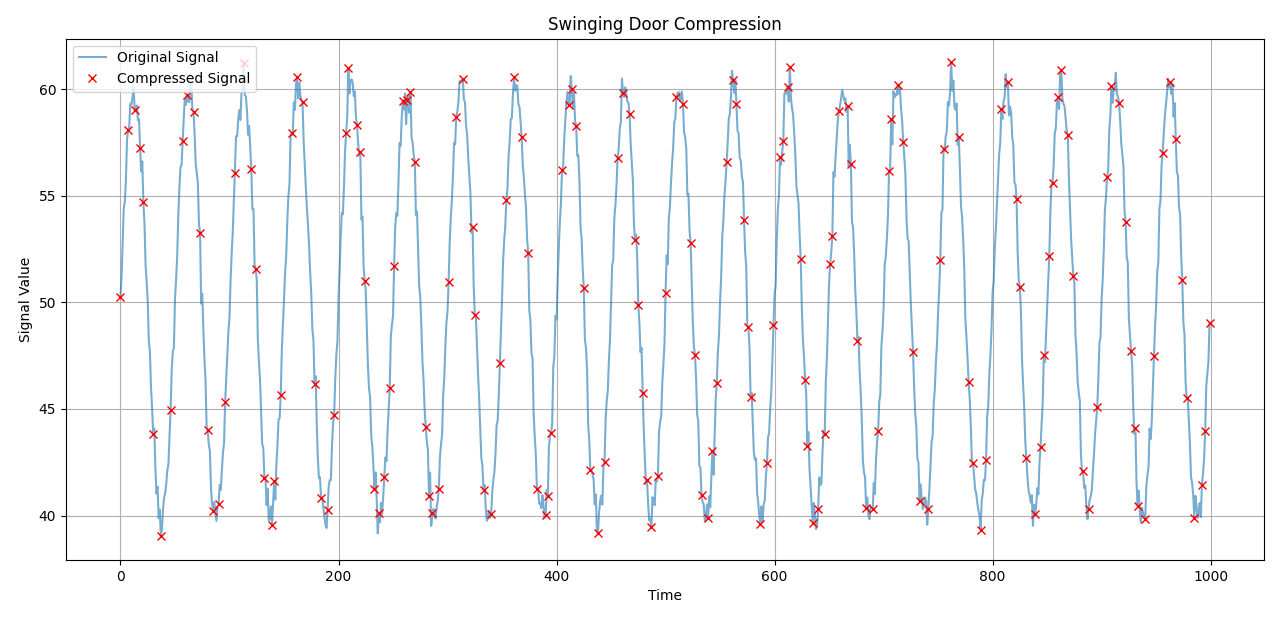}
    \includegraphics[width=\linewidth]{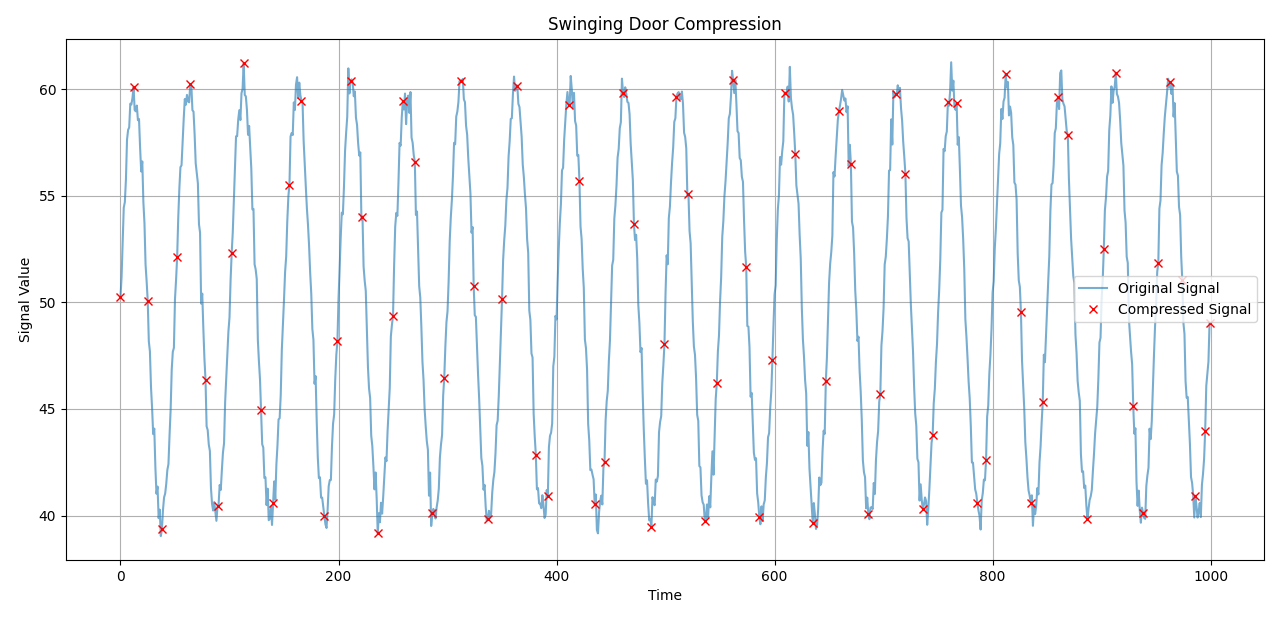}
    \caption{Compression of thresholds 0.713, 0.444, and 1.783}
\end{figure}

Applying a threshold-based compression algorithm (e.g., swinging door), data was reduced by approximately 60–65\%, which is consistent with industry norms (OSIsoft, 2020).

\ \\ 
\ \\ 
\ \\ 
\ \\ 
\ \\

\subsection{Effect on Data Quality and Signal Shape}

Compression can lead to subtle but important data distortions:

\begin{itemize}
	\item Mean deviation: ~2.1\%
	\item Standard deviation underestimation: ~8\%
    \item Missed transient anomalies: ~35\% when compression is too aggressive
\end{itemize}

Figure 2 shows how over-compression can impact anomaly detection recall.

\begin{figure}
    \centering
    \includegraphics[width=\linewidth]{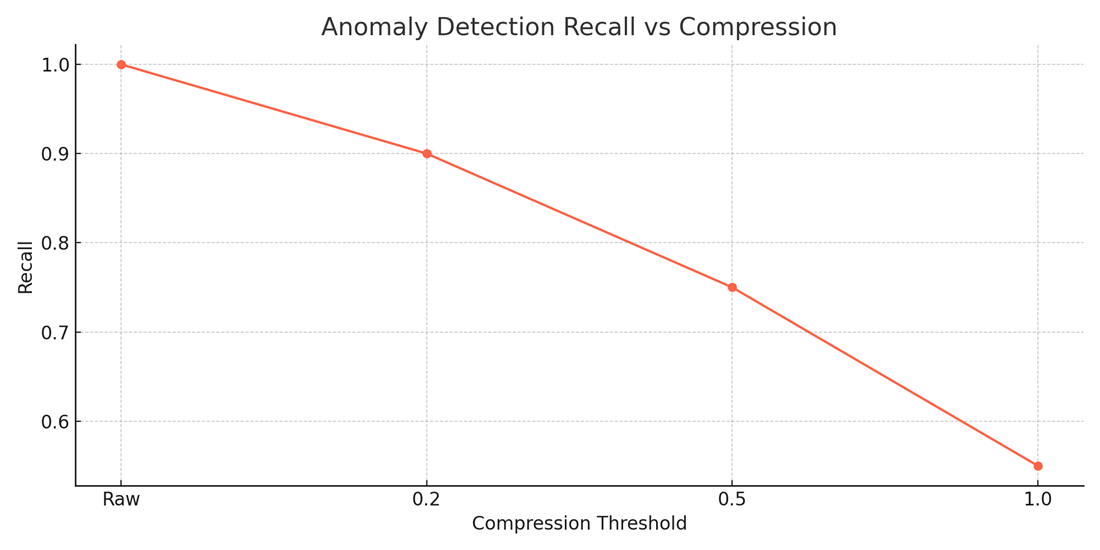}
    \caption{Anomaly Detection Recall vs Compression}
\end{figure}

\subsection{Impact on Univariate Models}

In this analysis, our primary goal was to conduct a comprehensive. To achieve this, we deliberately selected a diverse range of analytical techniques (univariate models), incorporating both traditional statistical models, such as AR, MA, ARMA, and Linear Regression (LR), and modern machine learning models, like LSTM, Random Forest (RF), and XGBoost.

This hybrid approach allows us to see if compression affects the type of model being used in a particular way. Statistical models are renowned for their interpretability, providing clear insights into the relationships between variables and the underlying structure of the data. In contrast, machine learning models often excel at predictive accuracy, as they can capture complex and non-linear patterns that simpler models might miss

\subsubsection{Baseline Performance (on Raw Data)}

Before looking at the effects of compression, let us see which models performed best on the original, raw data. The goal is to have the lowest Mean Absolute Error (MAE) and Mean Squared Error (MSE).

\begin{itemize}
	\item \textbf{Temperature Data}: The standout performers are Linear Regression (MAE: 0.2007, MSE: 0.0616), Random Forest (MAE: 0.2138, MSE: 0.0704), and ARMA (MAE: 0.2339, MSE: 0.0807). These models had significantly lower errors than the other models like MA and LSTM on this dataset.
    \\
	\item \textbf{Vibration Data}: The top models are ARMA (MAE: 0.3412, MSE: 0.1893) and Random Forest (MAE: 0.4044, MSE: 0.2705). Here again, traditional statistical and machine learning models outperformed the others in their raw state.
\end{itemize}

\subsubsection{The Impact of Data Compression}

Before looking at the effects of compression, let us see which models performed best on the original, raw data. The goal is to have the lowest Mean Absolute Error (MAE) and Mean Squared Error (MSE).

\begin{itemize}
	\item {\textbf{Temperature Data}: The effect of compression is highly model-dependent. Traditional time series models like Autoregressive (AR) and Moving Average (MA) show mixed results. The AR model benefits from compression, with an 11.15\% improvement in MAE and a substantial 114.44\% improvement in MSE. This suggests that compression may have removed some noise from the data, allowing the AR model to better capture the underlying patterns. In contrast, the MA model shows a negligible 0.50\% improvement in MAE and a significant 87.46\% degradation in MSE. This indicates that while the average error is slightly reduced, the compressed data introduces larger errors, which are heavily penalized by the MSE metric. The Autoregressive Moving Average (ARMA) model, a combination of AR and MA, performs exceptionally poorly with compressed data, showing a 66.91\% and 161.57\% degradation in MAE and MSE, respectively. This suggests that the compression technique used may have distorted the specific patterns that the ARMA model relies on.
    
    More complex models, such as Long Short-Term Memory (LSTM), also struggle with the compressed temperature data, with a 6.84\% and 32.73\% decrease in performance for MAE and MSE. LSTMs are a type of recurrent neural network that can learn long-term dependencies, and the information loss from compression appears to be detrimental to their performance. However, the most striking results are seen with the machine learning models. Linear Regression, Random Forest, and XGBoost all show remarkable improvements in MAE (381.29\%, 255.74\%, and 233.96\%, respectively). This indicates that for these models, the compressed data is significantly easier to predict. The compression might be acting as a form of feature engineering, simplifying the data and making the underlying relationships more apparent. However, it's important to note the corresponding degradation in MSE for these models (-94.55\%, -94.02\%, and -83.72\%), which points to the introduction of some large prediction errors. This dichotomy suggests that while the models are more accurate on average, they may be failing to predict certain data points accurately, a trade-off that would need to be considered in a practical application.}
    \end{itemize}
    
    \begin{table} 
	\centering 
	\begin{tabular}{C{0.2\linewidth} C{0.08\linewidth} C{0.08\linewidth} C{0.08\linewidth} C{0.08\linewidth}} 
        \cmidrule(r){2-5}
        
		& \multicolumn{2}{c}{Temp. Raw}  & \multicolumn{2}{c}{Vib. Raw}  \\
        \multicolumn{1}{l}{} & MAE & MSE & MAE & MSE\\
		\midrule
        AR & 3.08 & 11.87 & 1.55 & 3.38 \\
		MA & 3.16 & 12.42 & 3.19 & 12.84 \\
		ARMA & 0.23 & 0.08 & 0.34 & 0.18 \\
		LSTM & 2.95 & 13.51 & 4.35 & 27.02 \\
		LR & 0.20 & 0.06 & 0.46 & 0.33 \\
		RF & 0.21 & 0.07 & 0.40 & 0.27 \\
        XGBoost & 0.35 & 0.18 & 0.43 & 0.32 \\
		\bottomrule
	\end{tabular}
    \caption{Models' MSE and MAE Using Raw Data}

\end{table}

\begin{itemize}
	\item {\textbf{Vibration Data}: The impact of data compression on the vibration data models is generally less pronounced and more consistently negative than for the temperature data. The AR model shows a slight degradation in performance, with a -3.30\% change in MAE and a -6.16\% change in MSE. The MA model, on the other hand, sees a slight improvement, with a 1.64\% and 3.61\% improvement in MAE and MSE, respectively. This is a direct contrast to the temperature data, where the AR model benefited and the MA model did not. The ARMA model shows a mixed result, with a small 0.59\% improvement in MAE but a -3.07\% degradation in MSE.
    The more complex models continue to show a negative impact from compression. The LSTM model's performance degrades by -3.19\% for MAE and -4.91\% for MSE. The machine learning models, which saw huge improvements in MAE for the temperature data, do not fare as well with the vibration data. Linear Regression has a negligible 0.39\% improvement in MAE and 0.94\% in MSE. Random Forest and XGBoost, however, show a decline in performance, with -3.48\% and -9.84\% degradation in MAE and -3.70\% and -15.57\% in MSE, respectively. This suggests that the nature of the vibration data is different from the temperature data, and the compression technique used is not as beneficial, and in some cases, is detrimental. It's possible that the compression algorithm is removing crucial information from the vibration data that these models rely on to make accurate predictions.}
\end{itemize}

\begin{table} 
	
	\centering 
	\begin{tabular}{C{0.2\linewidth} C{0.08\linewidth} C{0.08\linewidth} C{0.08\linewidth} C{0.08\linewidth}} 
        \cmidrule(r){2-5}
        
		& \multicolumn{2}{c}{Temp. Comp.}  & \multicolumn{2}{c}{Vib. Comp.}  \\
        \multicolumn{1}{l}{} & MAE & MSE & MAE & MSE\\
		\midrule
        AR & 2.77 & 9.73 & 1.60 & 3.61 \\
		MA & 3.14 & 12.30 & 3.14 & 12.39 \\
		ARMA & 0.70 & 0.69 & 0.33 & 0.19 \\
		LSTM & 3.17 & 12.84 & 4.50 & 28.42 \\
		LR & 0.04 & 0.007 & 0.46 & 0.33 \\
		RF & 0.06 & 0.01 & 0.41 & 0.28 \\
        XGBoost & 0.10 & 0.0194 & 0.48 & 0.38 \\
		\bottomrule
	\end{tabular}
    \caption{Models' MSE and MAE Using Comp. Data}
\end{table}

\section{Discussion}

The results demonstrate that data compression must be aligned with the intended analytical purpose. Compression saves storage but risks analytical degradation if over-applied.

The effect of data compression on univariate model performance is not straightforward and depends heavily on the model, the nature of the data, and the compression technique used. For the temperature data, compression acted as a beneficial pre-processing step for several machine learning models, leading to significant improvements in MAE, although at the cost of increased MSE. For the vibration data, however, the impact of compression was mostly negative, suggesting that crucial information was lost during the process. These results highlight the importance of carefully evaluating the impact of data compression on a case-by-case basis. It is not a one-size-fits-all solution and can lead to either significant improvements or degradations in model performance. A thorough understanding of the data and the models being used is essential before applying any data compression techniques.

\begin{table} 
	\centering 
	\begin{tabular}{C{0.2\linewidth} C{0.12\linewidth} C{0.12\linewidth} C{0.12\linewidth} C{0.12\linewidth}} 
        \cmidrule(r){2-5}
        
		& \multicolumn{2}{c}{Temp. (\%)}  & \multicolumn{2}{c}{Vibr. (\%)}  \\
        \multicolumn{1}{l}{} & MAE & MSE & MAE & MSE\\
		\midrule
        AR & 11.2 & 114.4 & -3.3 & -6.2 \\
		MA & 0.5 & -87.5 & 1.6 & 3.6 \\
		ARMA & -67 & -161.5 & 0.6 & -3.1 \\
		LSTM & -6.8 & -32.7 & -3.2 & -4.9 \\
		LR & 381.2 & -94.6 & 0.3 & 0.94 \\
		RF & 255.7 & -94 & -3.5 & -3.7 \\
        XGBoost & 233.9 & -83.7 & -9.8 & -15.6 \\
		\bottomrule
	\end{tabular}
    \caption{MAE and MSE Improvements}
\end{table}

Key observations include:

\begin{itemize}
	\item Signal variability is a key determinant of safe compression limits.
	\item Shape-preserving compression (e.g., swinging door) is more reliable for analytics.
    \item Analytical models (especially anomaly detection and ML) require more conservative compression than simple reporting metrics.
\end{itemize}

\section{Conclusion}

This study provides a comprehensive analysis of the critical trade-off between data compression efficiency and analytical accuracy in modern industrial. As companies deal with huge and growing amounts of time-series data, compression has become an essential tool for managing storage costs and improving system performance. However, the research demonstrates that applying compression without considering the nature of the data and its intended analytical use can lead to significant degradation of data quality, ultimately undermining the very insights these systems are designed to generate.

The findings reveal that a "one-size-fits-all" approach to data compression is fundamentally flawed. Through simulations using the swinging-door algorithm, the paper establishes that the optimal compression strategy is highly dependent on the signal’s intrinsic characteristics. Smooth, slowly changing signals, such as temperature, can withstand aggressive compression with minimal loss of fidelity. In contrast, high-frequency, dynamic signals, like vibration data, require a more conservative approach to preserve the critical patterns and fluctuations necessary for accurate analysis. The paper highlights a direct, quantifiable relationship between compression levels and analytical integrity; as compression ratios increase, there is a corresponding rise in data reconstruction errors (RMSE) and a notable decline in the effectiveness of applications like anomaly detection.

Furthermore, the investigation into the impact on various univariate predictive models uncovered a complex and nuanced relationship. When some machine learning models were applied to smoother signals, compression unexpectedly improved Mean Absolute Error (MAE), suggesting it can act as a noise-reducing filter. However, for more volatile signals and for traditional time-series models like ARMA, compression was largely detrimental, indicating that essential predictive information was being discarded. The core conclusion is that compression strategies must be purpose-driven and adaptive. For applications where analytical precision is critical, such as predictive maintenance and anomaly detection, organizations must prioritize data integrity by implementing conservative, signal-aware compression thresholds.

\section{References}

[1] OSIsoft, "Exception, Compression, and their Impacts on PI System Performance," 2020. [Online]. Available: https://cdn.osisoft.com/osi/presentations/2023-AVEVA-San-Francisco/UC23NA-3PGK04-AVEVA\_Bregenzer\_Brent-Exception-Compression-and-their-Impacts-On-PI-System-Performance.pdf 

[2] J. Smith, "Lossless Compression of Wind Plant Data," IEEE Transactions on Industrial Informatics, vol. 9, no. 2, pp. 751-758, 2013. https://ieeexplore.ieee.org/document/6200401

[3] A. Brown, "Data Compression in Smart Distribution Systems via Singular Value Decomposition," IEEE Transactions on Industrial Electronics, vol. 61, no. 5, pp. 2314-2322, 2014. https://ieeexplore.ieee.org/document/7202904

[4] Y. Zhang, "The research of historical data compression and storage strategy in power dispatch SCADA system," ResearchGate, 2015. https://www.researchgate.net/publication/28774642\\
3\_The\_research\_of\_historical\_data\_compressi\\
on\_and\_storage\_strategy\_in\_power\_dispatch\_\\
SCADA\_system

[5] H. Lee, "Historical Multi-Station SCADA Data Compression of Distribution Management System Based on Tensor Tucker Decomposition," IEEE Transactions on Industrial Informatics, vol. 14, no. 4, pp. 1746-1755, 2018. https://ieeexplore.ieee.org/stamp/sta\\
mp.jsp?arnumber=8812699

[6] J. Kim, "Novel data compression technique for power waveforms using adaptive fuzzy logic," IEEE Transactions on Industrial Electronics, vol. 52, no. 3, pp. 834-841, 2005. https://ieeexplore.ieee.org/document/1458890

[7] Y. Chen, "Disturbance data compression of a power system using the Huffman coding approach with wavelet transform enhancement," IET Generation, Transmission \& Distribution, vol. 1, no. 3, pp. 431-438, 2007. https://digital-library.theiet.org/doi/10.1049/ip-gtd\%3A20030071

[8] H. Wang, "Real-Time Power-Quality Monitoring with Hybrid Sinusoidal and Lifting Wavelet Compression Algorithm," IEEE Transactions on Industrial Electronics, vol. 58, no. 3, pp. 1034-1041, 2011. https://ieeexplore.ieee.org/document/6236275

[9] J. Li, "A High Efficient Compression Method for Power Quality Applications," IEEE Transactions on Industrial Electronics, vol. 57, no. 3, pp. 1034-1041, 2010. https://ieeexplore.ieee.org/document/5738685

[10] Y. Zhang, "A Wavelet-Based Data Compression Technique for Smart Grid," IEEE Transactions on Industrial Informatics, vol. 8, no. 2, pp. 274-281, 2012. https://ieeexplore.ieee.org/document/5664816

[11] H. Lee, "Application of a Real-Time Data Compression and Adapted Protocol Technique for WAMS," IEEE Transactions on Industrial Electronics, vol. 62, no. 5, pp. 2834-2842, 2015. https://ieeexplore.ieee.org/document/6847242/refe\\
rences\#references

[12] A. Bagnall, J. Lines, A. Bostrom, J. Large, and E. Keogh, "The Great Time Series Classification Bake Off: A Review and Experimental Evaluation of Recent Algorithmic Advances," Data Min. Knowl. Discov., vol. 31, no. 3, pp. 606–660, 2017, doi: 10.1007/s10618-016-0483-9

\end{document}